\documentclass{article}
\usepackage{amsfonts, amsmath, amsthm, amssymb}
\usepackage[all]{xy}
\usepackage{amscd}

\newtheorem{theorem}{Theorem}[section]

\newtheorem{definitiontemp}[theorem]{Definition}
\theoremstyle{remark}

\newcommand{\fun}{\mathcal{O}}
\newcommand{\Z}{\mathbb{Z}}
\newcommand{\Proj}{\mathbb{P}}
\newcommand{\Q}{\mathbb{Q}}
\newcommand{\A}{\mathbb{A}}

\newcommand{\lb}{\Lambda_B}
\newcommand{\zp}{\mathbb{Z}_{\!p}}
\newcommand{\fp}{\mathbb{F}_{\!p}}

\begin{document}
\title{Supergeometry and Arithmetic Geometry.}
\author{A. Schwarz\thanks{Partly supported by NSF grant No. DMS 0505735.}
, I. Shapiro}
\date{}
\maketitle

\begin{abstract}
We define a superspace over a ring $R$ as a functor on a
subcategory of the category of supercommutative $R$-algebras.  As
an application the notion of a $p$-adic superspace is introduced
and used to give a transparent construction of the Frobenius map
on $p$-adic cohomology of a smooth projective variety over $\zp$
(the ring of $p$-adic integers).

\end{abstract}
\section {Introduction}

It is possible that all physical quantities are measured in
rational numbers (or even in integers if there exist elementary
length, elementary unit of time, etc). All other numbers should be
introduced only for mathematical convenience as it is much more
practical to work with an algebraically closed field that is
complete with respect to some norm.

Let us consider an oversimplified example where time and space are
quantized and the motion of a particle satisfies the equation
\begin{equation}
\Delta_2x(t)=F(x(t),t)
\end{equation}
Here $\Delta _2 $ stands for the second difference:
\begin{equation}
\Delta_2x(t)=[(x(t+2)-x(t+1))-(x(t+1)-x(t))]=x(t+2)-2x(t+1)+x(t)
\end{equation}
The equation (1) which is the finite difference analogue of the
Newton's second law, permits us to calculate recursively $x(t)$
for $t\in {\mathbb{Z}}$ if we know $x(0),x(1)$ and $F$; if the
initial data $x(0), x(1)$ and the ``force" $F$ are all integers
then all coordinates $x(t)$ will also be integers. In some sense,
all physical questions can be answered if we know only integer
numbers. However, if we would like to write down an explicit
solution of (1) even in the simplest situation when $F$ is a
linear function of $x(t)$ and does not depend explicitly on $t$ we
need irrational and complex numbers. Indeed, the general solution
to the equation
\begin{equation}
\Delta_2x(t)=ax(t)
\end{equation}
has the form
\begin{equation}
x(t)=A_1\lambda_1^t+A_2\lambda_2^t
\end{equation}
where $\lambda_1,\lambda_2$ are  solutions of a quadratic
equation.

Usually we take $\lambda _1,\lambda_2\in\mathbb{C}$; we are
working with the field of complex numbers that is algebraically
closed and complete with respect to the natural norm. However it
is possible to make another choice. It is well known that every
norm on ${\mathbb{Q}}$ is equivalent to either the standard norm
$\Vert x\Vert=\vert x\vert$ or $p$-adic norm
$$\Vert x \Vert _p=p^{-{\rm ord}_px}$$
where $p$ is a prime number. Here ${\rm ord}_px$ is equal to the
multiplicity of $p$ in the prime decomposition of $x$ if $x\in {
\mathbb{Z}}$; in general, ${\rm ord}_p(x/y)={\rm ord}_px-{\rm
ord}_py$. Completing ${\mathbb{Q}}$ with respect to the $p$-adic
norm we obtain the field of $p$-adic numbers ${\mathbb{Q}}_p$; the
completion of $\mathbb{Z}$ in the $p$-adic norm gives the ring of
$p$-adic integers $\mathbb{Z}_p$.   Since $\mathbb{Q}_p$ is not
algebraically closed, sometimes  it is convenient to work with
 the field $\mathbb{C} _p$  obtained as a completion of the
algebraic closure ${\bar{\mathbb{Q}}_p}$ of the field
$\mathbb{Q}_p$ with respect to an appropriate norm. The field
$\mathbb{C}_p$ is  algebraically closed and complete; in many ways
it is as convenient as  the field of complex numbers.

Of course, in most cases  it is better to work with the familiar
complex numbers, but situations do arise where $p$-adic methods
are more efficient. Let us suppose, for example, that the space
variable $x$ in (3) is periodic with period $N$, or, in other
words, $x$ takes values in  cyclic group ${\bf{Z}}_N$. (We are
using the notation  ${\bf{Z}}_p$ for $\mathbb{Z}/ p\mathbb{Z}$ to
distinguish it from the ring of $p$-adic integers $\mathbb{Z}_p$.)
Then the equation (3) splits into independent equations
corresponding to prime factors of $N$; it is natural to study the
elementary equations by means of $p$-adic methods.

There were numerous attempts to use $p$-adic numbers in quantum
field theory and string theory (see, for example, \cite {F} for
review of $p$-adic strings). It is clear that taking $p$-adic
physics seriously (i.e. considering it not as a formal gadget, but
as physical reality)  one should work with all prime numbers
simultaneously (in other words one should work with adeles). We
would like to emphasize that we  think that not only $p$-adic
numbers, but also irrational and complex numbers should be
regarded as a formal tool in the theory based on rational numbers.
If we accept this viewpoint there is no necessity to work in the
adelic setting.

In general, $p$-adic methods have a good chance to be useful when
interesting physical quantities are represented  by integer or
rational numbers. This is the case for topological sigma-models.
It was shown in \cite{inst}, \cite{instanton} that the instanton
numbers for sigma-models over complex numbers can be expressed in
terms of $p$-adic $B$-model, or, in other words, in terms of the
variation of Hodge structure on $p$-adic cohomology.  More
precisely, it was proven that instanton numbers can be expressed
in terms of the Frobenius map on $p$-adic cohomology and this fact
was used to analyze integrality of instanton numbers.

Recall that the Frobenius map of a field of characteristic $p$
transforms $x$ into $x^p$.  This map is an endomorphism of the
field, it is furthermore an automorphism for finite fields and an
identity map for the field $\mathbb{F}_p$ consisting of $p$
elements. Of course the $p$-th power map is an algebra
homomorphism for any $\fp$-algebra\footnote{An $R$-algebra is a
ring $A$ equipped with a ring homomorphism $R\rightarrow A$. This
notion is perhaps more familiar in the case when $R$ is a field.
All rings are assumed to be unital and associative.}. This last
observation leads to the Frobenius map on a variety over $\fp$.
One can consider the $p$-th power map also on the $p$-adic
numbers, but in this case it is not a homomorphism. We can  define
the Frobenius map on the $n$-dimensional affine or projective
space over $\zp$ by raising the coordinates to the $p$-th power.
However, a variety sitting inside (an affine or projective
variety)  will not,  in general, be invariant with respect to the
 Frobenius map.  Nevertheless,  one can define the Frobenius map
on the cohomology with coefficients in the ring of $p$-adic
integers $\mathbb{Z}_p$.

The standard approach to the construction is quite complicated; it
is based on the consideration of the so called  DP-neighborhood of
a variety that  has the same cohomology as the original variety
and at the same time is invariant with respect to the Frobenius
map. One of the main goals of the present  paper is to give a
simplified construction of the Frobenius map in terms of
supergeometry. The logic of our construction remains the same, but
the role of a DP-neighborhood is played by a $p$-adic superspace
having the same body as the original variety\footnote{Varieties we
consider are defined over the ring of $p$-adic integers
$\mathbb{Z}_p$. The elements of this ring can be represented as
formal power series $\sum a_kp^k$ where the coefficients $a_0,
a_1,  ....$ are integers between $0$ and $p-1$. If the series is
finite it specifies a conventional integer and the operations on
$p$-adic numbers coincide with operations in $\mathbb{Z}.$
Assigning the coefficient $a_0$ to a series $\sum a_kp^k$ we
obtain a homomorphism of $\mathbb{Z}_p$ onto the field
$\mathbb{F}_p$. Using this homomorphism we obtain from every
variety over the $p$-adic integers a variety over $\mathbb{F}_p$;
we say that this variety is the body of the variety over the
$p$-adic integers.}.

We hope that our approach to the Frobenius map will be much more
accessible to physicists than the standard one based on the
consideration of the crystalline site and DP-neighborhoods \cite
{bando}. Our considerations, as presented here, are not completely
rigorous; we did not want to exceed significantly the level of
rigor that is standard for a physics journal. However, we believe
that our results are of interest also to pure mathematicians and
therefore we have written a rigorous exposition of our construction of the Frobenius map
\cite{padic}.

It seems that arithmetic geometry should play an essential role in
string theory; the paper \cite{inst} sketches one way of applying
arithmetic geometry, but there are also other ways (let us mention
a series of papers by Schimmrigk, for example \cite{sch}).  The
present paper shows that, conversely, some ideas borrowed from
physics can be used to clarify some important notions  of
arithmetic geometry.

As we mentioned, our construction is based on the notion of a
$p$-adic superspace.  We hope that this notion will have also
other applications.  In particular, we expect that it can be used
to construct supersymmetric sigma-models and topological
sigma-models in the $p$-adic setting. The $p$-adic $B$-model used
in \cite{inst} was defined completely formally as the theory of
variations of Hodge structures. However it is natural to
conjecture that it can be defined in the framework of Lagrangian
formalism and that it is related to the (conjectural) $p$-adic
supersymmetric sigma-model.

One should notice that in an appropriate coordinate system the
operations in  the (super)group of supersymmetries are given by
polynomial formulas with integer coefficients; this means that one
can define the supersymmetry group over an arbitrary commutative
ring $R$ assuming that the coordinates belong to this ring (for
example, one may consider the supersymmetry group over
$\mathbb{Z}$; then we can talk about supersymmetry over integers).

To construct supersymmetric objects it is natural to introduce a
notion of superspace over ring $R$; this can be done in many ways.
In the conventional superspace over real numbers  coordinates are
considered as even or odd elements of a Grassmann algebra.  The
algebra of functions on the $(m,n)$-dimensional affine superspace
over $\mathbb{R}$ can be considered as a tensor product of the
algebra of smooth functions of $m$ real variables and the
Grassmann algebra with $n$ generators. In mathematical formulation
\cite{84} the conventional superspace is regarded as a functor on
the category of Grassmann algebras with values in the category of
sets (we assign to every  Grassmann algebra $\Lambda$ the set of
points with coordinates from $\Lambda$). A function on a
superspace is defined as a map of functors.  It is clear that the
functions on superspace described above specify  maps of functors
(natural transformations). It is not so clear that there are no
other natural transformations, but this can be proven \cite{vor}.
One can consider also superspaces where in addition to $m$ even
and $n$  odd coordinates we have $r$ even nilpotent coordinates;
then there is an additional factor  (the algebra of formal power
series with $r$ variables) in the tensor product specifying the
algebra of functions (see \cite{konech}).  It is important to
notice that even  considering only  superspaces over real numbers
one can modify the definition of a superspace in a variety of ways
to obtain different classes of functions. For example, we can
assume that the coordinates are even and odd elements of an
arbitrary supercommutative  algebra over $\mathbb{R}$; then only
polynomial functions are allowed.

If we take as the starting point the functorial approach, the
definition of a superspace over  a ring $R$ becomes very natural:
we should fix a subcategory of the category of  supercommutative
$R$-algebras and define a superspace as a functor  from this
subcategory to the category of sets.   \emph{The algebra of
functions depends on the choice of a subcategory.} Notice that the
functorial approach is common in algebraic geometry where a
variety over a ring $R$  specifies a functor on the category of
$R$-algebras with values in the category of sets. (A variety is
 singled out by polynomial equations with coefficients in
$R$ in affine or projective space over $R$. These equations make
sense if the coordinates are considered as elements of an
$R$-algebra; we can consider the corresponding set of solutions.
Similarly one can define the notion of a supervariety singled out
by means of polynomial equations in an affine or projective
superspace and construct the corresponding functor.)

We are interested in the case when $R$ is the ring of $p$-adic
integers $\mathbb{Z}_p$  and we define  the subcategory
$\Lambda(p)$ as the category of rings of the form $B\otimes
\Lambda$ where $B$ is a commutative ring such that $p^{N\gg 0}B=0$
and $B/pB$ does not contain nilpotent elements; $\Lambda$ stands
for a Grassmann ring.  By definition, a $p$-adic superspace is a
functor on the category $\Lambda(p)$ with values in the category
of sets. As always, we can define an affine superspace as a space
with $m$ even, $r$ even nilpotent and $n$ odd coordinates.
Functions on such a superspace are described in \cite{padic}. In
this paper we need only the case when $m=n=0$. If  $r=1$ (one even
nilpotent coordinate), then  functions have the form
$\sum\frac{c_n}{n!}z^n$ where $c_n\in\zp$.  The superspace at hand
can  be considered as an infinitesimal neighborhood $\widetilde
{pt}$ of a point on a line; it follows from the description of
functions that the cohomology of $\widetilde {pt}$ coincides  with
the cohomology of a point (this is a crucial observation that is
used in the construction of the Frobenius map).  The space
${\widetilde {pt}}^r$ with $r$ even nilpotent coordinates can be
regarded as an infinitesimal neighborhood of a point in the
$r$-dimensional affine space; its cohomology also coincides with
the cohomology of a point. Notice that the superspace $\widetilde
{pt}^r$ does not have any odd coordinates (even though Grassmann
algebras were used in its construction).

Let us restrict the functor defining a $p$-adic superspace to the
subcategory of  $\mathbb{F}_p$-algebras without nilpotent
elements. If there exists  a variety  over $\mathbb{F}_p$ that
specifies the same functor we say that this variety is the body of
the superspace at hand.

Consider a projective variety  $V$ over $\zp$. We mentioned
already that the Frobenius map acts on the projective space
sending a point  with homogeneous coordinates $(x_0:...:x_n)$ to
the point with coordinates $(x_0^p:....:x_n^p)$. The variety $V$
is not invariant with respect to the Frobenius map, but its body
is invariant. (We can define  the body  considering the functor
$X_V$ corresponding to the variety $V$ or we can use the simpler
definition in the footnote $^2$; both definitions lead to the same
result.) Now we consider the maximal subsuperspace
$\widetilde{X_V}$ of the projective space having the same body as
$V$ (and call it the infinitesimal neighborhood of $X_V$). It is
obvious that $\widetilde{X_V}$ is invariant with respect to the
Frobenius map. This means that the Frobenius map induces a
homomorphism on the cohomology groups of the $p$-adic superspace
$\widetilde{X_V}$. If $V$ is smooth one can prove that the
cohomology groups of $\widetilde{X_V}$ are isomorphic to the
cohomology groups of $V$. (The comparison of cohomology groups can
be reduced to local calculations and locally we use the
isomorphism between the cohomology of  $\widetilde {pt}^r$ and the
cohomology of a point.)  We thus obtain an action of Frobenius map
on cohomology groups  of a non-singular projective variety over
$\zp$ (Sec. 6).  In the last section  we demonstrate the
properties of Frobenius map that were used in \cite{inst}.  In the
paper \cite{fromir} the same properties were used to express the
Frobenius map on the cohomology of a Calabi-Yau threefold in terms
of the mirror map and instanton numbers.

\section{Superspaces}

Let us denote by $A$ a supercommutative ring (i. e. a
$\mathbb{Z}_2$-graded ring $A=A_0\oplus A_1$ where $a_0b=ba_0,
a_1a_1^{\prime}=-a_1^{\prime}a_1$ if $a_0\in A_0, a_1,
a_1^{\prime}\in A_1,  b\in A$).

One can say that $A^{p,q}=\{(x_1,...,x_p,\xi_1,...,\xi_q)|x_i\in
A_0,\xi_i\in A_1\}$ is a set of $A$-points of $(p,q)$-dimensional
superspace (of space with $p$ commuting and $q$ anticommuting
coordinates). One can consider $A^{p,q}$ as an abelian group or as
an $A_0$-module, but these structures will not be important in
what follows.

A parity preserving homomorphism $\varphi: A\rightarrow\tilde{A}$
induces a map $F(\varphi): A^{p,q} \rightarrow \tilde{A}^{p,q}$.
It is obvious that $F(\varphi\psi)=F(\varphi)\circ F(\psi)$ and
$F(id)=id$. This means that we can consider a functor $\A^{p,q}$
on the category of supercommutative rings taking values in the
category of sets and assigning $A^{p,q}$ to the ring $A$.

\emph{One can define a superspace as a functor from the category
of supercommutative rings into the category of sets}; then
$\A^{p,q}$ represents the $(p,q)$-dimensional affine superspace
(over $\Z$).

We define a map of superspaces as a map of functors. In particular
a map of  $(p,q)$-dimensional superspace into a $(p^{\prime},
q^{\prime})$-dimensional superspace is defined as a collection of
maps $\psi_A :  A^{p,q} \rightarrow A^{p^{\prime} , q^{\prime}}$
that are compatible with parity preserving homomorphisms of rings.
In other words, for every  parity preserving homomorphism $\varphi
: A\rightarrow \tilde{A}$ the  diagram
\[
\begin {CD}
     A^{p,q}           @>\psi _A >>         A^{p^{\prime}, q^{\prime}}   \\
     @V F(\varphi)VV             @V F(\varphi)VV \\
       \tilde{A}^{p,q}  @>   \psi_{\tilde{A}}>>  \tilde{A}^{p^{\prime},q^{\prime}}
\end {CD}
\]
should be commutative.

Let us denote by $\fun_{\Z}^{p,q}$ a $ \mathbb{Z} _2$-graded ring
of polynomials depending on $p$ commuting and $q$ anticommuting
variables and having integer coefficients. In other words:
$$  \fun_{\Z}^{p,q}=\mathbb{Z}[x^1, ... ,x^p]
\otimes \Lambda _{\mathbb{Z}}[\xi^1, ... , \xi^q]$$ is a tensor
product of the polynomial ring $\mathbb{Z}[x^1, ... ,x^p]$ and the
Grassmann ring $\Lambda_{\mathbb{Z}}[\xi^1, ... , \xi ^q]$, the
$\mathbb{Z}_2$-grading  comes from $ \mathbb{Z}_2$-grading of the
Grassmann ring.

It is easy to check that an  even element of $\fun_{\Z}^{p,q}$
determines a map of the $(p,q)$-dimensional affine superspace into
the $(1,0)$-dimensional affine superspace  and an odd element of
$\fun_{\Z}^{p,q}$ determines a map into $(0,1)$-dimensional affine
superspace. More generally, a row $(f_1, ... ,f_{p^{\prime}},
\varphi _1, ... ,\varphi _{q^{\prime}})$ of $p^{\prime}$ even
elements of $\fun_{\Z}^{p,q}$ and  $q^{\prime}$ odd elements of
$\fun_{\Z}^{p,q}$ specifies a map of $(p,q)$-dimensional
superspace into $(p^{\prime},q^{\prime})$-dimensional superspace
(a map of functors
$\mathbb{A}^{p,q}\rightarrow\mathbb{A}^{p^{\prime} ,
q^{\prime}}$). The construction is obvious: we substitute elements
of the ring $A$ instead of generators of $\fun_{\Z}^{p,q}$. One
can prove that all  maps of functors $\mathbb{A}^{p,q} \rightarrow
\mathbb{A}^{p^{\prime} , q^{\prime}}$ are described by means of
this construction.

More examples of superspaces can be constructed as affine
supervarieties. Affine supervariety (as  usual affine variety) can
be defined by means of polynomial equations:
     $$f_1(x^1, ... , x^p,\xi^1, ... , \xi^q)=0$$
     $$......$$
     $$f_{p^{\prime}}(x^1, ... , x^p,\xi^1, ... , \xi^q)=0$$
     $$\varphi_1(x^1, ... , x^p,\xi^1, ... , \xi^q)=0$$
     $$......$$
     $$\varphi_{q^{\prime}}(x^1, ... , x^p,\xi^1, ... , \xi^q)=0$$      More formally,
we consider a map of functors $F: \mathbb{A}^{p,q}\rightarrow
 \mathbb{A}^{p^{\prime} , q^{\prime}}$ corresponding to a row
     $(f_1, ... , f_{p^{\prime}},\varphi_1, ... , \varphi_{q^{\prime}})$
and define a set $\mathcal{B}_A$ of $A$-points of the supervariety
$\mathcal{B}$ as a preimage of zero by the map
$F(A):A^{p,q}\rightarrow A^{p^{\prime},q^{\prime}}$. It is obvious
that every parity preserving homomorphism $\varphi :A\rightarrow
\tilde{A}$ induces a map $\mathcal{B}_A \rightarrow \mathcal{B}_{
\tilde{A}}$.

Notice that to every affine supervariety we can assign an affine
variety (underlying variety) by dropping the equations $\varphi_i=0$ and setting
$\xi^i=0$ in the equations $f_i=0$, i.e. we neglect odd variables
and equations.

One can easily move from the affine to the projective picture as
in the usual commutative case.  Recall that the functor of points
corresponding to the projective space $\mathbb{P}^n$ can be
defined by the formula
$\mathbb{P}^n(A)=(A^{n+1})^{\times}/A^{\times}$, where
$(A^{n+1})^{\times}=\{(x_0,...,x_n)\in A^{n+1}|\sum A x_i=A\}$,
the group $A^{\times}$ of invertible elements of the ring $A$ acts
by means of componentwise multiplication. We can define the super
version of the projective space by setting
$\mathbb{P}^{n,m}(A)=\{(x_0,...,x_n,\xi_1,...,\xi_m)\in
(A_0^{n+1})^\times\times A_1^{m}\}/A_0^\times$ where $A$ is a
supercommutative ring. One easily checks that this results in a
functor thereby defining the projective superspace.

As before we may get a more general object, namely a projective
supervariety by considering the solutions of graded homogeneous
equations in the variables $x^i$ and $\xi^i$, that is solutions of
a system:

$$f_1(x^0, ... , x^n,\xi^1, ... , \xi^m)=0$$
     $$......$$
     $$f_{s}(x^0, ... , x^n,\xi^1, ... , \xi^m)=0$$
     $$\varphi_1(x^0, ... , x^n,\xi^1, ... , \xi^m)=0$$
     $$......$$
     $$\varphi_{t}(x^0, ... , x^n,\xi^1, ... , \xi^m)=0$$ where
$f_i\in\fun^{n+1,m}_{\mathbb{Z}}$ is an even homogeneous element
and $\varphi_i\in\fun^{n+1,m}_{\mathbb{Z}}$ is an odd homogeneous
element.

Again we can associate to the projective supervariety its
underlying variety: a projective variety obtained by neglecting
the odd variables and equations\footnote{In the terminology of
algebraic geometry a system of polynomial equations in affine or
projective space specifies an affine or projective scheme. The
term variety is reserved for schemes where local rings do not have
nilpotent elements; we will assume that  this condition is
satisfied. Similarly one should talk about affine superschemes
reserving the word supervariety for the case when underlying
scheme  is a variety.}.  The above constructions produce affine
and projective supervarieties as functors on the category of
supercommutative rings, hence they can be considered as
superspaces. A map of superspaces is a natural transformation of
the functors. Similarly, a function on a superspace is a natural
map  to the affine superspace $\mathbb{A}^{1,1}$.

There are some important modifications to these constructions that
will be useful to us. First of all instead of rings one can
consider algebras over a field or more generally $R$-algebras,
where $R$ is some fixed commutative ring. Modifying the source
category for our functors in this way leads to the notion of a
superspace over $R$. Furthermore, we can consider instead of the
category of all supercommutative $R$-algebras its various
subcategories.  It is worthwhile to note that the functions on a
superspace depend on the particular subcategory that we choose.
The most important example for us will be the subcategory of
supercommutative $R$-algebras that consists of rings of the type
$B\otimes\Lambda _{\mathbb{Z}}[\xi^1, ... , \xi^q]$ where $B$ is a
commutative $R$-algebra.  The concepts of affine and projective
supervarieties modify readily to this more general setting.

The construction of $\mathbb{A}^{p,q}$ can be generalized in the
following way.  Given an arbitrary ${\bf{Z}}_2$-graded $R$-module
$E=E_0\oplus E_1$, we can define a superspace over $R$, by setting
$E(A)=E_0\otimes A_0\oplus E_1\otimes A_1$.\footnote{A different
(in general) superspace can be obtained by considering
$E'(A)=\text{Hom}_R(E,A)$ where we require the morphisms to be
parity preserving.  When $E$ is free over $R$, this is the same as
the tensor construction.} When $R=\mathbb{Z}$ and $E$ is a free
abelian group we obtain $\mathbb{A}^{p,q}$.

An important example of a superspace that does not belong to the
types described above is defined by setting
$\widetilde{spt}(A)=A^{nilp}$ where $A^{nilp}$ is the subring of
nilpotent elements.  We will actually be more interested in
$\widetilde{pt}(A)=A_0^{nilp}$.  This is an example of a more
general and key construction to be discussed later.

Using the functorial approach to supergeometry it is easy to
define the notion of supergroup, super Lie algebra and their
actions.  Namely one should replace the target category of sets in
the definition of the superspace with the category of groups, Lie
algebras, etc.  $\A^{0,1}$ for instance can be given the structure
of a supergroup or that of a super Lie algebra.  A differential on
a $\mathbf{Z}_2$-graded $R$-module generates an action of
$\A^{0,1}$ on the corresponding superspace, conversely an action
of $\A^{0,1}$ gives a differential.  (Recall that a differential
on a $\mathbf{Z}_2$-graded $R$-module is a parity reversing
$R$-linear operator $d$, with $d^2=0$, satisfying the (graded)
 Leibnitz rule.)

\section{Body of a superspace}
Let us consider a subcategory $\mathcal{C}$ of the category of
supercommutative $R$-algebras having the property that for every
$A\in\mathcal{C}$ the quotient $A/A^{nilp}$ of the ring $A$ by the
ideal of nilpotent elements is also in $\mathcal{C}$.  The rings
$A/A^{nilp}$ where $A\in\mathcal{C}$ form a subcategory of the
category of commutative $R$-algebras. Note that by assumption, it
is a subcategory of $\mathcal{C}$ as well, denote it by
$\mathcal{C}_{red}$.

Recall that a superspace is a functor on $\mathcal{C}$.  Define
the body of the superspace to be its restriction to
$\mathcal{C}_{red}$.  We will always assume that the body of a
superspace is an algebraic variety.  This means that there exists
a variety over $R$ such that the restriction of the corresponding
functor on $R$-algebras to $\mathcal{C}_{red}$ coincides with the
restriction of the superspace to $\mathcal{C}_{red}$.  This
variety is not unique, however in the situation we consider later,
there exists such an ideal $\mathfrak{M}\subset R$ that
$\mathcal{C}_{red}$ can be identified with the subcategory of
commutative $R$-algebras consisting of $R/\mathfrak{M}$-rings
without nilpotent elements, thus the body is a unique
$R/\mathfrak{M}$-variety. If our superspace is an affine or
projective supervariety then the notion of body agrees with
the notion of underlying variety (if the body is considered as a
variety over $R$).

If $X$ is a superspace we will denote its body considered as a
variety by $X_{body}$.  Every subvariety $Z\subset X_{body}$
determines a subsuperspace $X|_Z$ of $X$ that can be thought of as
the largest subsuperspace of $X$ with body $Z$.  More precisely,
for every $A\in\mathcal{C}$ we have a map  $\pi_A:X(A)\rightarrow
X(A/A^{nilp})$ induced by the morphism $A\rightarrow A/A^{nilp}$.
The subvariety $Z$ specifies a subset $Z(A/A^{nilp})\subset
X(A/A^{nilp})$, we define $X|_Z(A)=\pi_A^{-1}(Z(A/A^{nilp}))$.

Consider an inclusion of superspaces $X\subset Y$, that is a
natural transformation $i$ such that for every $A$,
$i_A:X(A)\hookrightarrow Y(A)$ identifies $X(A)$ with a subset of
$Y(A)$. Note that the variety $X_{body}$ is a subvariety of
$Y_{body}$.  Define the infinitesimal neighborhood $\widetilde{X}$
of $X$ in $Y$, by setting $\widetilde{X}=Y|_{X_{body}}$. Clearly
$X\subset\widetilde{X}$.  For example, if $X$ is the origin in
$Y=\A^1$, the infinitesimal neighborhood of $X$ is the superspace
$\widetilde{pt}$ defined earlier.

\section{De Rham cohomology of a superspace}

Given a general superspace $X$, we associate to it a new
superspace $\Pi TX$ called the odd tangent space, defined by $\Pi
TX(A)=X(A\otimes\Lambda_\Z[\epsilon])$. It should be thought of as
the superspace parameterizing the maps from $\A^{0,1}$ to $X$.

To make the definition more transparent, consider the following
more familiar situation.  Let us assume that $X$ is an affine
algebraic supervariety defined by polynomial equations
$f_i(x^j,\xi^k)=0$ and $\varphi_i(x^j,\xi^k)=0$ as before.  Then
$\Pi TX$ is given by polynomial equations obtained by setting
$x^i=y^i+\epsilon\eta^i$ and $\xi^j=\zeta^j+\epsilon z^j$ in
$f_i=0$ and $\varphi_j=0$. (Here $y^i$, $z^j$ are even variables
and $\eta^i$, $\zeta^j$, $\epsilon$ are odd.)  $\Pi TX$ is then an
affine supervariety with coordinates $y^i$, $z^j$, $\eta^i$,
$\zeta^j$ constrained by the even equations $f_k(y^i,\zeta^j)=0$
and $\sum\eta^i\partial_{y^i}\varphi_k(y^i,\zeta^j)+\sum
z^j\partial_{\zeta^j} \varphi_k(y^i,\zeta^j)=0$ as well as the odd
equations $\varphi_k(y^i,\zeta^j)=0$ and
$\sum\eta^i\partial_{y^i}f_k(y^i,\zeta^j)+\sum
z^j\partial_{\zeta^j} f_k(y^i,\zeta^j)=0$.

The functions on $\Pi TX$ are (by definition) differential forms
on $X$.  The differential on the ring of functions on $\Pi TX$
(the ring $\Omega(X)$ of differential forms on $X$) can be defined
by $dy^i=\eta^i$, $d\eta^i=0$, $d\zeta^j=z^j$, and $dz^j=0$.

The differential on $\Pi TX$ has geometric origin that permits one
to define it for every superspace $X$.  Namely $\A^{0,1}$ has a
structure of a supergroup and therefore it acts on itself in an
obvious way.  Viewing $\Pi TX$ as the superspace parameterizing
maps from $\A^{0,1}$ to $X$ we obtain an action of $\A^{0,1}$ on
it.  Similarly we have an action of $\A^\times$ on $\Pi TX$
derived from the action of $\A^\times$ on $\A^{0,1}$ via
$\A^\times(A)\times\A^{0,1}(A)\rightarrow\A^{0,1}(A)$ which is
just the multiplication map $A^\times\times A_1\rightarrow A_1$.
The actions of $\A^{0,1}$ and $\A^\times$ on $\Pi TX$ are
compatible in the sense that the semi-direct product
$\mathbb{A}^{0,1}\rtimes\mathbb{A}^\times$ acts on $\Pi TX$.
Furthermore this construction is functorial, i.e. a natural
transformation $\psi:X\rightarrow Y$ induces a natural
transformation between the odd tangent spaces $d\psi:\Pi
TX\rightarrow \Pi TY$ that is compatible with the action of
$\mathbb{A}^{0,1}\rtimes\mathbb{A}^\times$.

The action of $\mathbb{A}^{0,1}$ on $\Pi TX$ and thus on the
differential forms on $X$ (= functions on $\Pi TX$) specifies a
differential $d$ on $\Omega(X)$ that can be used to define the
cohomology groups $\text{ker}d/\text{im}d$.  These groups are in
fact $R$-modules. The action of $\A^\times$ descends to the
subquotient and gives a grading on the cohomology.

In the case when $X$ is a smooth affine algebraic variety
(considered as a functor) and the source category is the full
category of supercommutative $R$-algebras, the cohomology groups
coincide with the familiar de Rham cohomology.  They may or may
not coincide with them if the source category is different.

To define the de Rham cohomology in a non-affine (but still smooth
case\footnote{In the non-smooth case one follows the procedure
below applied to $Y|_X$ instead of $X$ itself, where $Y$ is
``smooth" and contains $X$. For this to work one has to choose an
appropriate source category for the functors specifying the
superspaces.}) recall that we assumed that the body of the
superspace $X$ is an algebraic variety.  For every open subvariety
$U\subset X_{body}$, consider the restriction $X|_U$ of $X$ to
$U$.  Define $\Omega^{\bullet}_{X/R}(U)$ as the $R$-algebra of
differential forms on $X|_U$, i.e. functions on $\Pi T(X|_U)$.  As
observed above $\Omega^{\bullet}_{X/R}(U)$ is a differential
$R$-algebra with the differential given by the action of
$\A^{0,1}$ and the grading given by the action of $\A^{\times}$.
The collection of $\Omega^{\bullet}_{X/R}(U)$ for every $U$
specify a sheaf of differential $R$-modules on $X_{body}$.  We
define the de Rham cohomology of $X$ as hypercohomology of this
sheaf.  That is we consider the $R$-module of \v{C}ech cochains
with values in $\Omega^{\bullet}_{X/R}$ and define the
differential on these as the sum of the differential on
$\Omega^{\bullet}_{X/R}(U)$ and the \v{C}ech differential.
Corresponding cohomology groups are by definition the de Rham
cohomology of $X$.  If $X$ is a smooth projective variety this
definition will produce cohomology groups isomorphic to the
familiar de Rham cohomology groups for a choice of the source
category that will interest us.

It is easy to see from the definitions that the De Rham cohomology
is a contravariant functor from the category of superspaces to the
category of graded $R$-algebras.  Thus any endomorphism of $X$
induces an endomorphism on the cohomology.

\section{Cohomology of an infinitesimal neighborhood.}
As we mentioned already an infinitesimal neighborhood of a point
$pt$ in the affine line $\A^1$ is the superspace $\widetilde{pt}$
given by a functor that assigns to a supercommutative ring $A$,
the nilpotent part $A_0^{nilp}$ of $A_0$.  If the source category
is the category of all supercommutative $\Q$-algebras, then the
ring of functions on $\widetilde{pt}$ is easily seen to be the
ring of formal power series over $\Q$, namely $\Q[[z]]=\{\sum a_n
z^n\}$.  The ring of differential forms is then
$\Q[[z]][dz]=\{\sum a_n z^n +\sum b_n z^n dz\}$.  (Here $z$ is
even and $dz$ odd as usual.)  The differential transforms $\sum
a_n z^n +\sum b_n z^n dz$ into $\sum n a_n z^{n-1} dz$.  We see
immediately that any element of $\text{ker}d$ belongs to
$\text{im}d$ except $a_0$, thus the map on cohomology induced by
the embedding of $pt$ into $\widetilde{pt}$ is an isomorphism.

It was important to work with $\Q$-algebras in the above
considerations. If we instead consider the category of all
supercommutative rings then we should replace $\Q[[z]]$ and
$\Q[[z]][dz]$ with $\Z[[z]]$ and $\Z[[z]][dz]$ respectively.  It
is clear that $\widetilde{pt}$ now has non-trivial higher
cohomology.\footnote{Naturally because we can no longer divide by
$n$.}

However one can define a subcategory of the category of
supercommutative rings in such a way that the cohomology of $pt$
and $\widetilde{pt}$ are naturally isomorphic.  Namely we should
use the category $\Lambda(p)$ consisting of objects
$\lb=B\otimes\Lambda _{\mathbb{Z}}[\xi^1, ... , \xi^q]$ where $B$
is a commutative ring such that $p^{N\gg 0} B=0$ and $B/pB$ has no
nilpotent elements.\footnote{Examples of such rings $B$ are
${\bf{Z}}_{p^k}$ and ${\bf{Z}}_{p^k}[x]$} Here $p$ is a fixed
prime (sometimes we need to assume that $p>2$).  A superspace with
the source category $\Lambda(p)$ will be called a \emph{$p$-adic
superspace}. In what follows we work with $p$-adic superspaces.

The functions on $\widetilde{pt}$ can then be identified with
formal series $\sum a_n z^n /n!$ and the forms on $\widetilde{pt}$
with $\sum a_n z^n /n!+\sum b_n z^n/n! dz$ where $a_n,b_n\in\zp$.

It is not difficult to prove that every series of this kind
specifies a function (or a form).  We notice first that
$\lb^{nilp}=pB+\lb^+$, where we denote by $\lb^+$ the ideal
generated by $\xi_i$s.  Then the claim above follows from these
observations:

\emph{A}.  For every nilpotent element $\zeta\in\Lambda_{\Z}$ we
can consider $\zeta^n/n!$ as an element of $\Lambda_\Z$,
furthermore $\zeta^n/n!=0$ for $n$ large, thus $\sum a_n z^n /n!$
can be evaluated at $\zeta$.

\emph{B}.  The ring $\lb$ can be considered as a $\zp$-algebra
since multiplication by an infinite series $\sum a_n p^n$ makes
sense since we have $p^{N\gg 0} B=0$.

\emph{C}.  For every $\eta\in pB$ the expression $\eta^n/n!$ makes
sense as an element of $B$ since $p^n/n!$ in the reduced form has
no $p$ factors in the denominator.  Furthermore, if $p>2$ then
$\sum a_n z^n /n!$ can be evaluated at $p$ to obtain an element of
$\zp$ and so $\sum a_n z^n /n!$ can be evaluated at $\eta$.

The statement that these are all the functions on $\widetilde{pt}$
is more complicated.  The proof is given in \cite{padic}, Sec.
4.1. From the viewpoint of a physicist this proof is irrelevant:
we can restrict ourselves to the functions described above and
not worry about other functions.

The differential on forms on $\widetilde{pt}$ is given by the
formula $d(\sum a_n z^n/n!+b_n z^n/n! dz)=\sum a_n z^{n-1}/(n-1)!
dz$.  It follows immediately from this formula that  the
cohomology of $\widetilde{pt}$ coincides with that of $pt$.

Let us now consider the  cohomology of the infinitesimal
neighborhood $\widetilde{X}$ of a smooth variety $X$ singled out
by a single equation $f(z)=0$. Locally the infinitesimal
neighborhood is a direct product of $X$ with $\widetilde{pt}$;
hence its cohomology coincides with the cohomology of $X$.  More
precisely we have a local expression for a differential form of
degree $s$ on $\widetilde{X}$: $w=\sum a_n f^n/n!+\sum b_n f^n/n!
df$, where $a_n$ and $b_n$ are differential forms on $X$ of degree
$s$ and $s-1$ respectively. One can derive the fact we need  from
the formula
$$h(w)=(-1)^{s-1}\sum_{n=0}^{\infty}b_n f^{n+1}/(n+1)!.$$ This
expression  establishes a homotopy equivalence between sheaves of
differential forms on $X$ and $\widetilde{X}$.

The above statements are particular cases of a general theorem
valid for a smooth subvariety $X$ of a smooth variety $Y$.  (More
generally, $X$ and $Y$ can be supervarieties.)  That is: \emph{If
$X$ and $Y$ are considered as superspaces over the category
$\Lambda(p)$ or over the category of all supercommutative
$\Q$-algebras then the cohomology of the infinitesimal
neighborhood $\widetilde{X}$ of $X$ in $Y$ is naturally isomorphic
to the cohomology of $X$.}\footnote{Recall that we defined
cohomology as hypercohomology of the complex of sheaves.  The
embedding of $X$ into $\widetilde{X}$ induces a map of the
corresponding complexes of sheaves that one can show, through
local analysis, is a quasi-isomorphism by constructing an explicit
homotopy. In the case when $X$ has codimension $1$ this homotopy
was constructed above.}

It follows from the above theorem that the cohomology of an
infinitesimal neighborhood of a smooth $X$ does not depend on
which particular embedding into a smooth $Y$ one
chooses.\footnote{It seems one can prove that this remains true
even in the case of a singular $X$.  This prompts a definition of
the cohomology of a singular $X$ as the cohomology of its
infinitesimal neighborhood in some smooth $Y$.  In the case of a
$p$-adic superspace $X$ obtained from a variety over $\fp$ we
should obtain the crystalline cohomology of the variety as the
cohomology of $\widetilde{X}$.}

\section{Frobenius map on the $p$-adic cohomology.}
Let V be a smooth projective variety over the ring $\zp$ of
$p$-adic integers. There is a standard way to define its De Rham
cohomology $H^\bullet(V;\zp)$ with coefficients in $\zp$.  One can
define the action of the Frobenius morphism on $H^\bullet(V;\zp)$.
The usual way to obtain this action is based on crystalline
cohomology. We will demonstrate a simpler construction in terms of
supergeometry.

Let us consider $V$ as a functor on the category $\Lambda(p)$ and
denote it by $X_V$.\footnote{$V$ being a variety over $\zp$
defines a functor from commutative $\zp$-algebras to sets.  For
any $\lb\in\Lambda(p)$ define $X_V(\lb)=V(\lb^{even})$.  Since
$\lb^{even}$ is a commutative $\zp$-algebra, this is well defined.
 We note that $V$ considered as a functor on $\Lambda(p)$ is not the same thing,
 for a general $V$, as $V$ considered as a variety, rather it
 is the $p$-adic completion.  However it is known that
the De Rham cohomology of $V$ is isomorphic to the De Rham
cohomology of $X_V$ in the smooth projective case.} We noticed
previously that a ring
 belonging to $\Lambda(p)$ can
be considered as a $\zp$-algebra, thus the de Rham cohomology of
$X_V$ as defined previously yields a graded $\zp$-algebra.  As
mentioned previously, the de Rham cohomology of $X_V$ coincides
with the de Rham cohomology of its infinitesimal neighborhood
$\widetilde{X_V}$ in the projective space $\Proj^n$ (considered as
a functor on $\Lambda(p)$).  Note that while $X_V$ is obtained
from a variety $V$, there is no variety that yields
$\widetilde{X_V}$.  The body of $\widetilde{X_V}$ coincides with
the body of $X_V$ and is a variety over the field $\fp$. (The body
is defined as a functor on the category $\Lambda(p)_{red}$ of
rings of the form $B/pB$ without nilpotent elements.  This is
exactly the category of $\fp$-algebras without nilpotent
elements.)

The Frobenius map on the projective space over $\zp$ transforms a
point with homogeneous coordinates $(x_0:...:x_n)$ into  the point
with homogeneous coordinates $(x_0^p:...:x_n^p)$, this is clearly
a natural transformation on the $p$-adic superspace $\Proj^n$. The
subfunctor $X_V$ is not preserved by this map, but its
infinitesimal neighborhood $\widetilde{X_V}$ is invariant.  This
follows immediately from the  definition of $\widetilde{X_V}$ as
the maximal subsuperspace with the same body as $X_V$ since the
body of $X_V$ is invariant inside the body of $\Proj^n$ under the
Frobenius map.\footnote{The invariance of the body of $X_V$ inside
the body of $\Proj^n$ is most easily explained by pointing out
that the Frobenius map on $\Proj^n$ as defined above has a more
natural (equivalent) definition when restricted to the subcategory
of $\Lambda(p)$ consisting of $\fp$-algebras (i.e. we consider
only those $B$ with $pB=0$).  It can be constructed by observing
that since the $p^{\text{th}}$ power map is a homomorphism of
$\fp$-algebras (as $a^p=a$ for $a\in\fp$) it induces an
endomorphism of any functor on the subcategory of $\fp$-algebras.
 Because the body is defined as the restriction of the functor to the
subcategory $\Lambda(p)_{red}$ which contains only $\fp$-algebras,
this constructs an action of Frobenius on the body of any $p$-adic
superspace that evidently preserves any subbody.}

Now the action of the Frobenius on $\widetilde{X_V}$ induces an
action on the de Rham cohomology of $\widetilde{X_V}$ which is the
same as the de Rham cohomology of $X_V$ which in the smooth
projective case is the same as the usual de Rham cohomology of
$V$.  We thereby obtain the celebrated lifting of Frobenius to
characteristic $0$.\footnote{As a bonus we see that the cohomology
of $V$ depends only on its restriction to $\fp$.}

\section{Hodge filtration}

The grading on the sheaf of differential forms on a variety $V$
determines a descending filtration $F^k$ (namely $F^k$ consists of
forms of degree at least $k$).  The filtration on forms specifies
a filtration on the cohomology, called the Hodge filtration,
denoted by $F^k$ as well.  To analyze the behavior of the
Frobenius map with  respect to the Hodge filtration on
$H^\bullet(V;\zp)$, where $V$ is a smooth projective variety, one
should define a filtration $\widetilde{F}^k$ on the sheaf of
differential forms on the infinitesimal neighborhood
$\widetilde{V}$ of $V$ in such a way that the isomorphism between
$H^\bullet(V;\zp)$ and $H^\bullet(\widetilde{V};\zp)$ identifies
the two filtrations.  (The exposition given here is inspired by
\cite{mazur}.)

Let us assume for simplicity that $V$ can be singled out locally
by one equation $f=0$; as we have mentioned already, there is the
following local expression for a differential form on
$\widetilde{V}$: $w=\sum a_n f^n/n!+\sum b_n f^n/n! df$, where
$a_n$ and $b_n$ are differential forms on $V$.  We say that $w$
above belongs to $\widetilde{F}^k$ if $a_n\in F^{k-n}$ and $b_n\in
F^{k-n-1}$. The filtration $\widetilde{F}^k$ has the desired
property.\footnote{This is due to the fact that this filtration is
preserved by the homotopy that establishes the isomorphism of
cohomologies.}

Unfortunately the Frobenius does not preserve the Hodge
filtration, however certain $p$-divisibility conditions are
satisfied.  These divisibility estimates were crucial in the proof
of integrality of instanton numbers \cite{inst,instanton}.

It is easy to check that if $a$ is a local function then
$Fr(a)=a^p+pb$ where $b$ is some other local function, thus
$Fr(f^k/k! da_1 ...da_s)=(f^p+pg)^k/k!
d(a_1^p+pb_1)...d(a_s^p+pb_s)$ is divisible by $p^k/k!\cdot p^s$.
Using this one can show that
$$Fr(F^k)\subset p^k H^\bullet (V;\zp),\,\text{if}\,
p>\text{dim}V.$$ This follows from the estimate $Fr(F^k)\subset
p^{[k]} H^\bullet (V;\zp)$ where $[k]$ is maximal such that
$p^{[k]}$ divides all numbers $p^n/n!$ with $n\geq k$. If $k<p$
then $p^{[k]}=p^k$ and  in the case when $k>\text{dim}V$, $F^k$ is
trivial, thus one can assume that $k\leq \text{dim}V$.

\vspace{\baselineskip} \noindent\textbf{Acknowledgements.} We are
deeply indebted to V. Vologodsky for  his help with understanding
of the standard approach to the construction of Frobenius map and
to M. Kontsevich  and A. Ogus  for interesting discussions.

\noindent Department of Mathematics, University of California,
Davis, CA, USA \newline \emph{E-mail address}:
\textbf{schwarz@math.ucdavis.edu}
\newline \emph{E-mail address}:
\textbf{ishapiro@math.ucdavis.edu}

\end{document}